\documentclass[twocolumn]{aastex62}

\usepackage{amsmath}	% Advanced maths commands
\usepackage{amssymb}	% Extra maths symbols

\newcommand{\be}{\begin{equation}}
\newcommand{\ee}{\end{equation}}

\graphicspath{{./}{figures/}}

\shorttitle{A concordance model for FRB~121102}
\shortauthors{Margalit \& Metzger}
\begin{document}

\title{A concordance picture of FRB~121102 as a flaring magnetar embedded in a magnetized ion-electron wind nebula}

\email{btm2134@columbia.edu}

\author{Ben Margalit}
\altaffiliation{NASA Einstein Fellow}
\affiliation{Department of Physics and Columbia Astrophysics Laboratory, Columbia University, New York, NY 10027, USA}
\affiliation{Astronomy Department and Theoretical Astrophysics Center, University of California, Berkeley, Berkeley, CA 94720, USA}

\author{Brian D.~Metzger}
\affiliation{Department of Physics and Columbia Astrophysics Laboratory, Columbia University, New York, NY 10027, USA}

%% Note that the \and command from previous versions of AASTeX is now
%% depreciated in this version as it is no longer necessary. AASTeX 
%% automatically takes care of all commas and "and"s between authors names.

%% AASTeX 6.2 has the new \collaboration and \nocollaboration commands to
%% provide the collaboration status of a group of authors. These commands 
%% can be used either before or after the list of corresponding authors. The
%% argument for \collaboration is the collaboration identifier. Authors are
%% encouraged to surround collaboration identifiers with ()s. The 
%% \nocollaboration command takes no argument and exists to indicate that
%% the nearby authors are not part of surrounding collaborations.

%% Mark off the abstract in the ``abstract'' environment. 
\begin{abstract}

The fast radio burst FRB 121102 has repeated multiple times, enabling the identification of its host galaxy and of a spatially-coincident, compact, steady (`persistent') radio synchrotron source.  It was proposed that FRB 121102 is powered by a young flaring magnetar, embedded within a decades-old supernova remnant.  Using a time-dependent one-zone model, we show that a single expanding magnetized electron-ion nebula (created by the same outbursts likely responsible for the FRBs) can explain all the basic properties of the persistent source (size, flux, self-absorption constraints) and the large but decreasing rotation measure (RM) of the bursts.  The persistent emission is powered by relativistic thermal electrons heated at the termination shock of the magnetar wind, while the RM originates from non-relativistic electrons injected earlier in the nebula's evolution and cooled through expansion and radiative losses.  The model contains few free parameters, which are tightly constrained by observations: the total energy injected into the nebula over its history, $\sim 10^{50}-10^{51}$ erg, agrees with the magnetic energy of a millisecond magnetar; the baryon loading of the magnetar outflow (driven by intermittent flares) is close to the neutron star escape speed; the predicted source age $\sim 10-40$ years is consistent with other constraints on the nebula size.  
For an energy input rate $\dot{E} \propto t^{-\alpha}$ following the onset of magnetar activity, we predict secular decay of the RM and persistent source flux, which approximately follow ${\rm RM} \propto t^{-(6+\alpha)/2}$ and $F_{\nu} \propto t^{-(\alpha^2+7\alpha-2)/4}$, respectively. 

\end{abstract}

%% Keywords should appear after the \end{abstract} command. 
%% See the online documentation for the full list of available subject
%% keywords and the rules for their use.

%% From the front matter, we move on to the body of the paper.
%% Sections are demarcated by \section and \subsection, respectively.
%% Observe the use of the LaTeX \label
%% command after the \subsection to give a symbolic KEY to the
%% subsection for cross-referencing in a \ref command.
%% You can use LaTeX's \ref and \label commands to keep track of
%% cross-references to sections, equations, tables, and figures.
%% That way, if you change the order of any elements, LaTeX will
%% automatically renumber them.
%%
%% We recommend that authors also use the natbib \citep
%% and \citet commands to identify citations.  The citations are
%% tied to the reference list via symbolic KEYs. The KEY corresponds
%% to the KEY in the \bibitem in the reference list below. 

\section{Introduction}

Fast radio burts (FRB) are short pulses of coherent radio emission lasting less than a few milliseconds \citep{Lorimer+07,Keane+12, Thornton+13, Spitler+14, Ravi+15, Petroff+16, Champion+16,Lawrence+17} with large dispersion measures (DM $\approx 300-2000 \,{\rm pc\, cm^{-3}}$), well above the contribution from the Milky Way and thus implicating an extragalactic origin.  The cosmological distance of at least one FRB was confirmed by the discovery of a repeating FRB~121102 \citep{Spitler+14,Spitler+16} and its subsequent localization \citep{Chatterjee+17} to a dwarf star-forming galaxy at a redshift of $z=0.1927$ \citep{tbc+.2017}.  Radio interferometric localization of FRB~1211012 revealed a compact (size $<$ 0.7 pc) luminous ($\nu L_{\nu} \sim 10^{39}$~erg\,s$^{-1}$) steady radio synchrotron source coincident to within $\lesssim 40 \, {\rm pc}$ of the FRB location \citep{mph+.2017}.  Another important clue to FRB~121102 comes from its enormous rotation measure, RM $\sim 10^{5}$ rad m$^{-2}$ (\citealt{Michilli+18}; see also \citealt{Masui+15}), which greatly exceeds those of other known astrophysical sources, with the exception of 
Sgr A* and
the flaring magnetar SGR J1745-2900 located in the Galactic Center \citep{Eatough+13}.  

Though dozens of models have been proposed for FRBs, most are ruled out by a repeating, cosmological source like FRB~121102.  Among the few surviving possibilities are bursts created from a young flaring magnetar \citep{Popov&Postnov13,Lyubarsky14,Kulkarni+14,Katz16,Lu&Kumar16,Metzger+17,Nicholl+17c,Kumar+17,Beloborodov17,Lu&Kumar17}.   
Supporting this connection are the atypical properties of the host galaxy of FRB~121102, particularly its small size and high specific star formation rate \citep{Bassa+17}, which are similar to those which preferentially host long gamma-ray bursts and superluminous supernovae \citep{Metzger+17}, transient events independently attributed to magnetar birth (e.g.~\citealt{Duncan&Thompson92,Thompson+04,Kasen&Bildsten10}).  In such a model, the spatially-coincident persistent radio source could be understood as emission from a compact magnetized nebula surrounding the young (decades to centuries old) neutron star, embedded behind the expanding supernova ejecta shell \citep{Murase+16,Metzger+17,Kashiyama&Murase17,Omand+18}.  The nebula is powered by nearly continual energy release from the magnetar, likely during the same sporadic flaring events responsible for the repeated radio bursts \citep{Beloborodov17}.   

While no single piece of evidence supporting the magnetar model for FRB~121102 is alone convincing, in aggregate the weight of evidence becomes more compelling.  In $\S\ref{sec:magnetar}$, we briefly summarize the physical model and current observational constraints.  In $\S\ref{sec:model}$ we present a one-zone model for an expanding magnetized electron-ion wind nebula surrounding the young flaring neutron star.  For physically-motivated parameters, we show that the properties of FRB~121102 and its persistent source are quantitatively consistent with the magnetar model.  Based on this surprisingly tightly constrained `concordance picture', we make predictions for the future evolution of the source properties.    

\section{Observational Constraints}
\label{sec:magnetar}

The main reservoir responsible for powering both the FRB and persistent radio emission is the magnetic energy of the magnetar,
\be
E_{\rm B_{\star}} \simeq B_{\star}^{2}R_{\star}^{3}/6 \approx 3\times 10^{49} B_{16}^{2}\, {\rm erg}\,,
\label{eq:EB}
\ee
where $B_{\star} = B_{16}\times 10^{16}$ G is the interior magnetic field strength and $R_{\star} \simeq 12$ km is the neutron star radius.  Rotational energy may be important at early times, particularly in the case of the magnetars born with millisecond rotation periods, as needed to power gamma-ray burst jets or luminous supernovae; however, given the rapid rate at which magnetars undergo magnetic braking, rotational energy is less significant for decades-old sources of interest and is thus neglected hereafter.  

\begin{figure}
\centering
\includegraphics[width=0.5\textwidth]{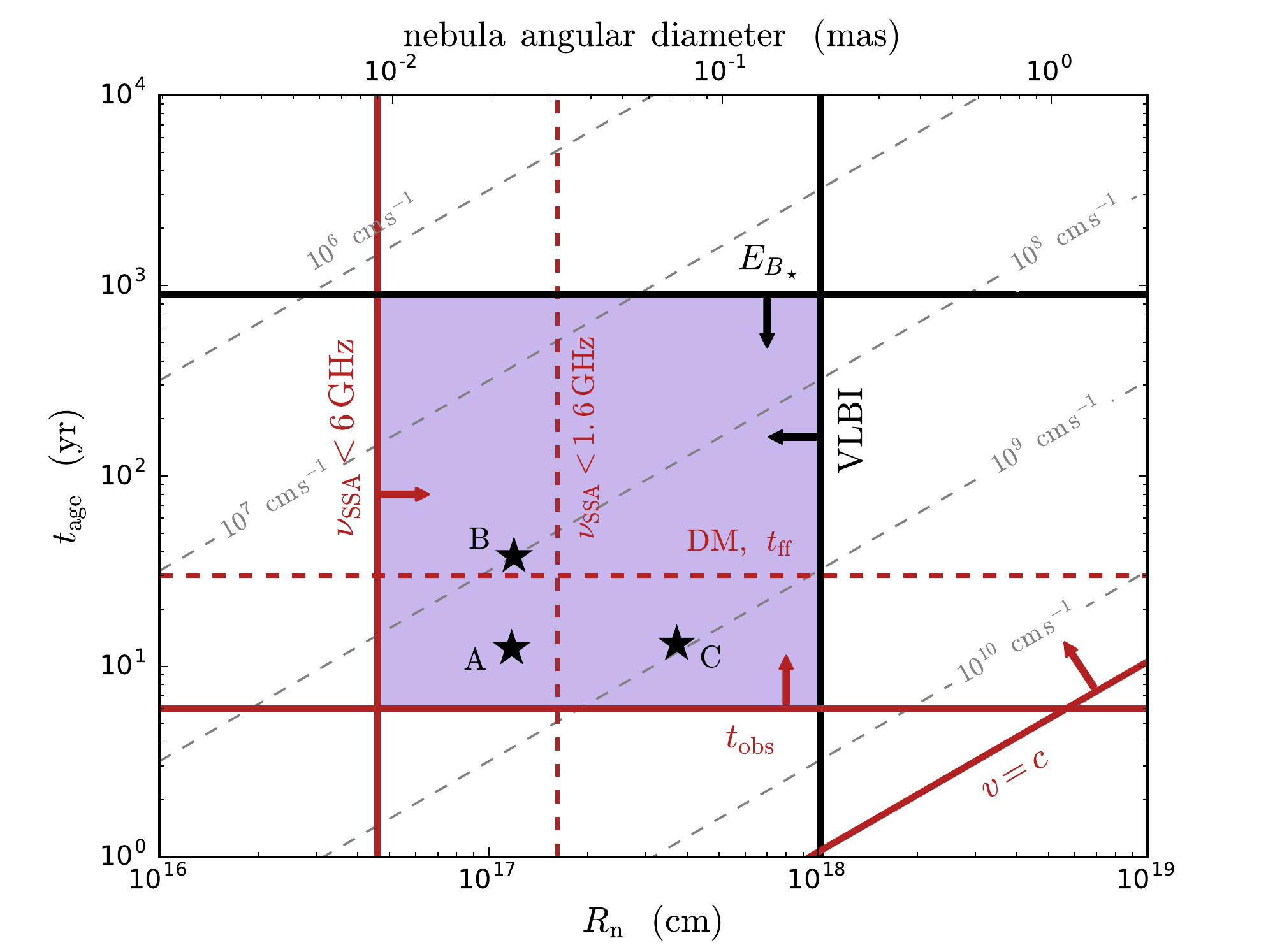}
\caption{Observational constraints on the age, $t_{\rm age}$, of the magnetar giving rise to FRB~121102 and the radius $R_{\rm n}$ of the compact synchrotron nebula, are shown.  These include: lower limits on $R_{\rm n}$ based on the lack of self-absorption features in the synchrotron spectrum (eq.~\ref{eq:R17_constraint_ssa}, shown for both $1.4$ GHz and $6$ GHz); upper limits on $R_{\rm n}$ from VLBI imaging; upper limits on the source age from the magnetic energy budget (eq.~\ref{eq:TB} for $B_{\star} \sim 10^{16}$ G); and approximate lower limits on the source age based on the supernova ejecta not overproducing the DM or free-free optical depth.  The shaded blue region is the allowed parameter space.  Dashed grey curves show the implied expansion velocity of the nebula, $R_{\rm n}/t_{\rm age}$.  Our best fit-models, which fit the persistent source flux and RM, are shown as stars.  The timescale over which magnetic energy is released from the magnetar interior, $t_{\rm mag}$ (eq.~\ref{eq:tmag}), which may be associated with the FRB active lifetime \citep{Beloborodov17}, also falls within the concordance region.}
\label{fig:timescales}
\end{figure}

The injection of particle and magnetic energy by the active magnetar inflates a compact synchrotron nebula behind the expanding supernova ejecta shell.  Fig.~\ref{fig:timescales} summarizes constraints that can be placed on the age of FRB~121102 since its birthing supernova, $t_{\rm age}$, and the radius of the synchrotron radio nebula, $R_{\rm n}$ 
\citep{Metzger+17,Waxman17}.  A lower limit of $t_{\rm age} \gtrsim 6$ yrs follows from the currently active time since its discovery.  An upper limit is set by the requirement of powering the persistent source luminosity, $\nu L_{\nu} \sim 10^{39}$ erg s$^{-1}$, over a timescale $\sim t_{\rm age}$, 
\be
t_{\rm age} \lesssim \frac{E_{\rm B_{\star}}}{\nu L_{\nu}} \approx 900\,B_{16}^{2}\,{\rm yr},
\label{eq:TB}
\ee
under the conservative assumption that the radio-emitting electrons are fast-cooling close to the present epoch.  

An additional age limit of $t_{\rm age} \gtrsim 10-100$ yr follows from the requirement that the supernova ejecta be transparent at $\sim$ 1 GHz to free-free absorption, and to not overproduce constraints on the time derivative of the dispersion measure (\citealt{Connor+16,Piro16,Margalit+18}).  However, the precise value of this lower limit depends on the free-free optical depth of the ejecta, which in turn depends on the level of photo-ionization by the spin-down powered nebula \citep{Margalit+18}.  The latter will be smaller for magnetars with strong dipole magnetic fields $B_{\rm d} \gtrsim 10^{15}$ G, due to their rapid magnetic braking.  As a final consistency check, \citet{Nicholl+17} show that if all FRB sources repeat in a manner similar to FRB~121102, then the birth rate of FRB-producing sources is consistent with those of superluminous supernovae or long GRBs for magnetar active life times of $t_{\rm age} \sim 60\xi^{-1}(\eta/0.1)^{-1}$ yr, where $\xi<1$ is the duty cycle and $\eta<1 $ is the FRB beaming fraction (cf.~\citealt{Lu&Kumar17,Law+17}).

The radius of the magnetar-inflated nebula $R_{\rm n}$ must be smaller than that of the freely-expanding supernova ejecta shell, $R_{\rm ej} \approx v_{\rm ej}t_{\rm age} \approx 0.1 (t_{\rm age}/10\,{\rm yr}) \, {\rm pc}$, where $v_{\rm ej} \sim 10^{4}$ km s$^{-1}$ is the typical mean ejecta speed for hydrogen-poor supernovae.  An upper limit on the nebula diameter $2R_{\rm n} \lesssim 0.66$ pc follows from VLBI imaging \citep{Marcote+17}.  A lower limit on the nebula size follows from the lack of a clear signature of synchrotron self-absorption (SSA) above $6 \, {\rm GHz}$
in the spectral energy distribution (eq.~\ref{eq:R17_constraint_ssa}); uncertainty in this constraint follows from the fact that the spectrum does show a break, possibly attributed to SSA, between 1.6 and 6 GHz (see top panel of Fig.~\ref{fig:luminosity}).

Combining the above constraints, the allowed parameter space, $t_{\rm age} \sim 10-1000$ yr and $R_{\rm n} \sim 10^{17}-10^{18}$ cm (Fig.~\ref{fig:timescales}), is consistent with a nebula of size $R_{\rm n} \sim 0.1-1 R_{\rm ej}$ for $v_{\rm ej} = 10^{9}$ cm s$^{-1}$.  The age also broadly agrees with theoretical predictions for the active lifetimes of magnetars.  Magnetic flux escapes from the interior of the young magnetar on a timescale set by ambipolar diffusion from the star's neutron core \citep{Beloborodov&Li16},
\be
t_{\rm mag} \sim 400 L_{\rm km}^{1.6}B_{16}^{-1.2}\,{\rm yr},
\label{eq:tmag}
\ee
where $\delta B_{\star} \sim B_{\star}/2$ and $L_{\delta B} = L_{\rm km}$ km are the amplitude and length-scale of the magnetic field fluctuations.  For $L_{\rm km} \sim 1-10$ allowed by the neutron star size, the predicted lifetimes range from decades to centuries for strong-field $B_{16} \gtrsim 1$ millisecond magnetars (like those we find are responsible for FRB 121102), to $t_{\rm age} \gtrsim 10^{3}-10^{4}$ yr for weaker-field magnetars, as may characterize those in our Galaxy.

The large rotation measure RM $\sim 10^{5}$ rad m$^{-2}$ of FRB~121102 shows that it is embedded in a dense electron-ion plasma (\citealt{Michilli+18}; see also \citealt{Masui+15}).  The magnetic field of the medium responsible for the RM must exceed $\sim 1$ mG \citep{Michilli+18}.  Though too high for the ISM of the host galaxy, such a large field strength might instead be attributed to the persistent synchrotron nebula.  The plasma nebula composition must be ion-electron, rather than electron-positron, because the RM contribution of the latter case is zero \citep{Michilli+18}.  Although this disfavors a pulsar-like wind nebula dominated by rotational energy input (e.g.~\citealt{Murase+16,Metzger+17,Kashiyama&Murase17}), we will show that it follows naturally for a nebula powered by magnetic energy \citep{Beloborodov17}.  The energetic plasma outflow that accompanied the giant flare of SGR~1806-20 in 2004 \citep{Palmer+05} was indeed inferred to be heavily mass-loaded from its sub-relativistic expansion speed and radio afterglow (\citealt{Granot+06}).  

\begin{figure}
\centering
\includegraphics[width=0.45\textwidth]{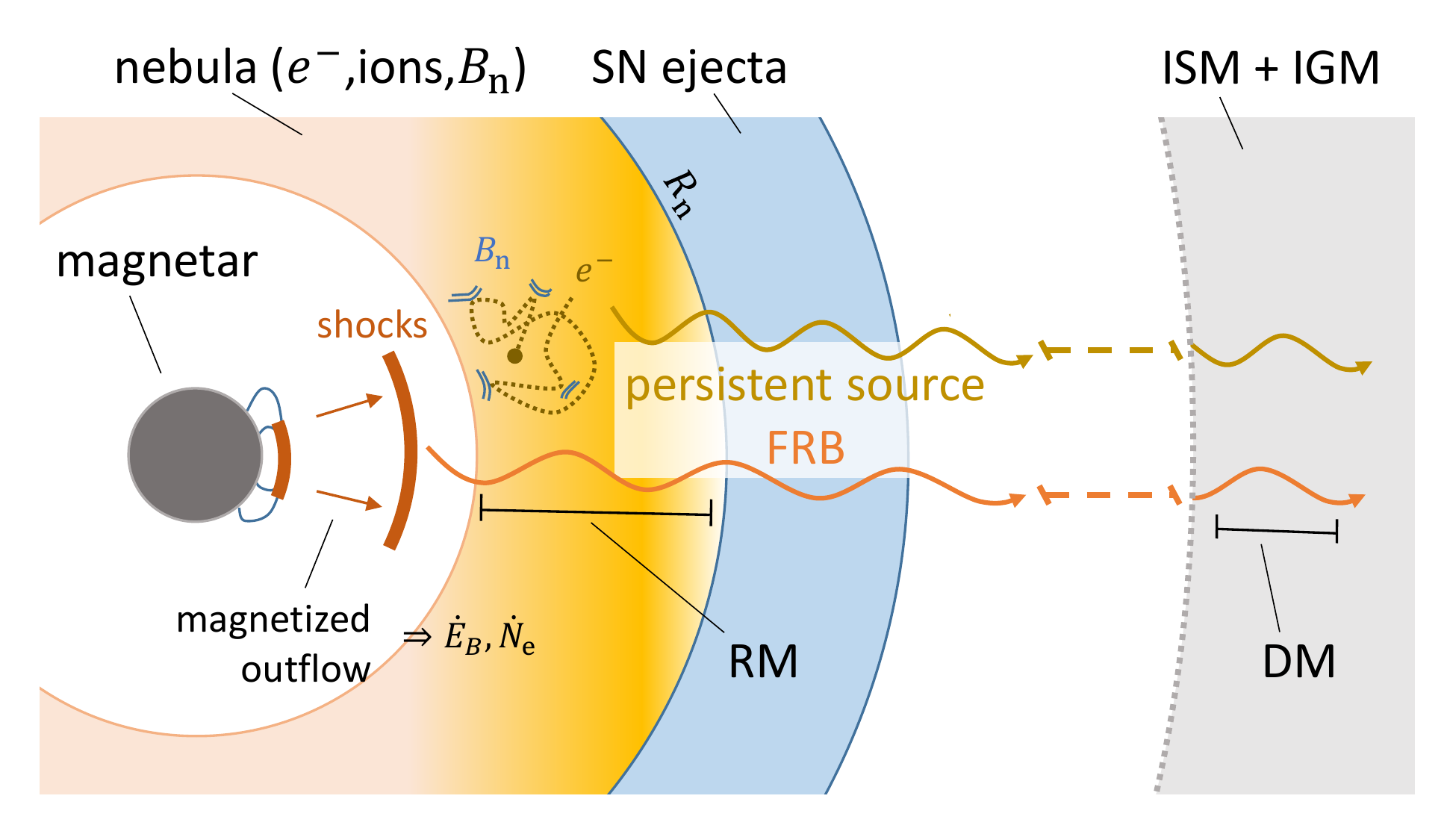}
\caption{Schematic picture of the concordance model: energetic magnetar flares episodically eject magnetized baryon-loaded outflows. The outflows terminate at the base of a nebula inflated by previous magnetar flares and/or rotationally powered winds, injecting particles, $\dot{N}_e$, and magnetic energy, $\dot{E}_B$, into the nebula. Energetic electrons gyrating within this magnetized nebula emit synchrotron radiation, observed as the `persistent' radio source associated with FRB~121102, while lower Lorentz factor electrons within the same nebula Faraday rotate FRB pulses (originating interior to the nebula) producing the large observed RM. Both signals propagate through the SN ejecta, which at current epochs must be free-free transparent and contribute negligibly to the DM that is accumulated on larger scales.}
\label{fig:cartoon}
\end{figure}

The RM of FRB~121102 was furthermore observed to decline by $\sim$10\% over a 7 month interval \citep{Michilli+18}.  This may suggest that a turbulent magnetized environment surrounds the burst, as in the Galactic Center magnetar \citep{Eatough+13}.  Alternatively, the decline may implicate secular evolution originating from the source being embedded in an expanding, diluting magnetized medium, either from the supernova shock wave interacting with dense circumstellar gas (\citealt{Piro&Gaensler18}) or the same burst-powered synchrotron nebula responsible for the quiescsent radio emission \citep{Margalit+18}.  
Figure~\ref{fig:cartoon} shows a schematic diagram of the latter possibility, which we now explore in greater depth.
%We now explore the latter possibility in greater depth.

\section{One-Zone Nebula Model}
\label{sec:model}

%We consider a freely-expanding ejecta shell of mass $M_{\rm ej}$ and velocity $v_{\rm ej}$, with a mean radius $R_{\rm ej} \approx v_{\rm ej}t$.  The free-expansion assumption is valid provided we are at early times compared to the Sedov-Taylor deceleration time of the shell interacting with the surrounding circumburster medium.  In the case of gamma-ray burst supernovae or SLSNe, typical values are $M_{\rm ej} \approx 3-20M_{\odot}$ and $v_{\rm ej} \approx 10^{4}$ km s$^{-1}$,
%implying a Sedov-Taylor time of $t_{\rm dec} \approx 300-600 n_0^{-1/3}$ \, {\rm yr} for ISM densities $n = n_0 \, {\rm cm}^{-3}$.

We assume that the nebula is spherical and freely expanding, with a characteristic radius $R_{\rm n} = v_{\rm n}t$ at time $t$ after the explosion and a constant radial velocity $v_{\rm n} \lesssim v_{\rm ej} \sim 10^{4}$ km s$^{-1}$. The magnetar releases its free magnetic energy $E_{B_{\star}}$ (eq.~\ref{eq:EB}) into the nebula at a rate that we model as a power-law in time
\be \label{eq:Edot}
\dot{E} = (\alpha-1) \frac{E_{B_{\star}}}{t_{0}}\left(\frac{t}{t_{0}}\right)^{-\alpha}, ~~~~~ t \ge t_{0}, ~ \alpha > 1
\ee
where $t_{0}$ is the onset of its active period, which may be controlled by the timescale for magnetic flux to begin to leak out of the magnetar core (eq.~\ref{eq:tmag}).  We model $\dot{E}$ as a smooth function of time, but in reality the energy release could occur intermittently, e.g. associated with the same discrete flaring events responsible for powering the repeating FRB \citep{Beloborodov17}.  Although the nebula may not strictly expand with a constant velocity, the ratio of the nebula radius to that of the freely-expanding supernova ejecta, $R_{\rm ej} \sim v_{\rm ej}t$, is expected to be a weakly varying function of time \citep{Chevalier77}, which for characteristic nebula energies obeys $R_{\rm n} \sim 0.1-1R_{\rm ej}$.

The magnetar injects ions, and their associated electrons, into the nebula at the rate
\be \label{eq:Ndot_e}
\dot{N}_{\rm e} = (1+\sigma)^{-1} \chi^{-1}\dot{E},
\ee
where 
$\sigma$ is the magnetization of the injected outflow, and $\chi$ the mean energy per particle and we have assumed a proton-electron composition.  Even if the initial phase of a magnetar flare produces an ultra-relativistic electromagnetic pulse with $\chi \gg m_{\rm p} c^{2}$, $\sigma \gg 1$  (e.g.~\citealt{Lyubarsky14,Beloborodov17}), the ejecta containing the bulk of the energy may not be so ``clean".  Indeed, values $\chi \approx 0.16-10$ GeV are motivated by the baryon loading of the outflow associated with the giant flare of SGR~1806-20 \citep{Palmer+05,Granot+06} and the minimum value set by the requirement to escape the gravitational potential of the neutron star, $\chi_{\rm min} = GM_{\star}m_{\rm p}/R_{\star} \sim 0.2$ GeV, where $M_{\star} = 1.4M_{\odot}$ is the neutron star mass.  We adopt a fiducial value of $\sigma = 0.1$, as the efficiency of synchrotron maser emission is sharply peaked about this upstream magnetization (e.g.~\citealt{hoshino_92}).  \citet{Beloborodov17} proposed that FRB emission is produced by internal shocks as ultra-relativistic flare ejecta collides with the slower, baryon-loaded wind created by the accumulation of previous flares.

Energy injected by the mildly relativistic magnetar wind thermalizes at a termination shock before entering the nebula.  Although some of the power of a collisionless magnetized shock is transferred into the ions, we assume that the electrons are also efficiently heated, e.g. by small-scale turbulence generated by the ions ahead of the shock \citep{Sironi&Spitkovsky11} or other plasma scale processes.  Electrons are thus heated upon entering the nebula to a mean particle Lorentz factor,
\be
\bar{\gamma} \approx \frac{\chi}{2 m_{\rm e} c^{2}} \approx 196\left(\frac{\chi}{\chi_{\rm min}}\right) .
\label{eq:gammai}
\ee

The number density of electrons with Lorentz factors between $\gamma$ and $\gamma + d\gamma$ in the nebula is defined as $N_\gamma d\gamma$,
%The Lorentz factor distribution of electrons in the nebula $N_{\gamma}$ is defined such that $n_{\rm e} = \int N_\gamma d\gamma$ where $n_{\rm e}$ is the electron number density, 
and obeys a continuity equation
\be
\frac{\partial}{\partial t}N_{\gamma} + \frac{\partial}{\partial \gamma}\left(\dot{\gamma} N_{\gamma}\right) 
- 3\frac{\dot{R}_{\rm n}}{R_{\rm n}}N_\gamma
%- \frac{\partial}{\partial \gamma} \left( E(\gamma) \gamma^2 \frac{\partial}{\partial \gamma} \left(\frac{N_\gamma}{\gamma^2}\right) \right) 
= \dot{N}_{\gamma} .
\ee
The source term is assumed to be a relativistic Maxwellian of temperature $kT = \bar{\gamma} m_e c^{2} / 3 \approx \chi / 6$ (eq.~\ref{eq:gammai}),
%\be
%\dot{N}_{\gamma} \left( \gamma \right) = \frac{\dot{N}_{\rm e}}{4\pi R_{\rm n}^3/3} \frac{ 3 / \bar{\gamma}  }{  K_2(3/\bar{\gamma}) } \gamma^2 \beta e^{-3\gamma/\bar{\gamma}}
%\ee
%where 
normalized such that
the total injection rate $\dot{N}_{\rm e} = V_{\rm n} \int \dot{N}_\gamma d\gamma$,
where $V_{\rm n}$ is the nebula volume,
is determined by eq.~(\ref{eq:Ndot_e}),
%$\beta = \sqrt{1-\gamma^{-2}}$, 
and the loss term
%\be
$\dot{\gamma} = \dot{\gamma}_{\rm adiab} + \dot{\gamma}_{\rm syn} + \dot{\gamma}_{\rm IC} + \dot{\gamma}_{\rm brem}$,
%\ee
includes adiabatic expansion (Vurm \& Metzger 2017)
\begin{equation}
\dot{\gamma}_{\rm adiab} = -\frac{1}{3}\gamma \beta^{2}\frac{d \ln V_{\rm n}}{dt} 
= - \gamma \beta^2 \frac{\dot{R}_{\rm n}}{R_{\rm n}} ,
%\frac{1}{3}\gamma \beta^{2}\frac{d \ln n_{\rm e}}{dt} 
%= \frac{1}{3}\gamma \beta^{2}\left(\frac{\dot{N}_{\rm b}}{N_{\rm b}} - 3\frac{\dot{R_{\rm n}}}{R_{\rm n}}\right) 
%\nonumber \\
%&=& -\frac{1}{3}\gamma \beta^{2}\left(-\frac{\dot{N}_{\rm b}}{N_{\rm b}} + \frac{3}{5t_{0}}(6-\alpha)\right)
\end{equation}
synchrotron and inverse-Compton radiation
\be \label{eq:gammadot_syn}
\dot{\gamma}_{\rm syn, IC} = -\frac{4}{3}\frac{\sigma_{\rm T}}{m_{\rm e} c} \beta^{2}\gamma^{2} 
\begin{cases}
%\frac{B_{\rm n}^2}{8\pi} f_{\rm ssa}(\gamma) ,
f_{\rm ssa}(\gamma) B_{\rm n}^2 / 8\pi &, \, {\rm syn}
\\
%\frac{L_{\rm rad}}{4\pi c R_{\rm n}^2}
L_{\rm rad} / 4\pi c R_{\rm n}^2 &, \, {\rm IC}
\end{cases}
\ee
and bremsstrahlung
\begin{equation}
\dot{\gamma}_{\rm brem} \approx -\frac{5}{3} c \sigma_{\rm T} \alpha_{\rm fs} n_{\rm e} \gamma^{1.2},
\end{equation}
%and inverse-Compton
%\begin{equation}
%\dot{\gamma}_{\rm IC} = -\frac{4}{3}\frac{\sigma_{\rm T}}{m_e c} \beta^{2}\gamma^{2} \frac{L_{\rm rad}}{4\pi c R_{\rm n}^2},
%\end{equation}
where $\alpha_{\rm fs} \simeq 1/137$ and $\sigma_{\rm T} = 6.65\times 10^{-25} \, {\rm cm}^{2}$.

The magnetic field in the nebula, $B_{\rm n} = (6 E_{\rm B}/R_{\rm n}^{3})^{1/2}$, is related to its total magnetic energy, $E_B$.  The latter is governed by injection of magnetic energy from the source and adiabatic losses,
\begin{equation}
\frac{dE_B}{dt} = -\frac{\dot{R}_{\rm n}}{R_{\rm n}} E_B + \frac{\sigma}{1+\sigma} \dot{E},
\end{equation}
where we have assumed that the field is tangled and thus evolves as a gas with an effective adiabatic index $\Gamma = 4/3$.  

At each time we calculate the synchrotron luminosity,
\begin{equation}
L_\nu = 4\pi^2 R_{\rm n}^2 \frac{j_\nu}{\alpha_\nu} \left( 1 - e^{-\alpha_\nu R_{\rm n}} \right) ,
\end{equation}
using the emissivity and absorption coefficient
\begin{equation}
j_\nu = \int \frac{N_\gamma P_\nu(\gamma)}{4\pi}  d\gamma
, ~ 
\alpha_\nu = - \int \frac{\gamma^2 P_\nu(\gamma)}{8\pi m_{\rm e} \nu^2} \frac{\partial}{\partial \gamma} \left[ \frac{N_\gamma}{\gamma^2} \right] d\gamma ,
\end{equation}
%\begin{equation}
%j_\nu = \frac{1}{4\pi} \int N_\gamma P_\nu(\gamma)  d\gamma ,
%\end{equation}
%\begin{equation}
%\alpha_\nu = - \frac{1}{8\pi m_e \nu^2} \int \gamma^2 P_\nu(\gamma) \frac{\partial}{\partial \gamma} \left[ \frac{N_\gamma}{\gamma^2} \right] d\gamma ,
%\end{equation}
respectively, and where
\begin{equation}
P_\nu = \frac{2 e^3 B_{\rm n}}{\sqrt{3} m_{\rm e} c^2}F\left(\frac{\nu}{\nu_{\rm c}}\right) , ~ F(x) \equiv x \int_{x}^{\infty} K_{5/3}(y) dy
\end{equation}
is the spectral power of a synchrotron emitting electron with characteristic frequency $\nu_{\rm c} = \gamma^2 eB / 2\pi m_{\rm e} c $ \citep[e.g.][]{Rybicki&Lightman},
%\begin{equation} \label{eq:nuc}
%\nu_{\rm c} = \frac{\gamma^2 eB }{ 2\pi m_e c } .
%\end{equation}

Synchrotron self-absorption
is important in our regimes of interest, and affects the emitted spectrum as well as
the synchrotron cooling rate of relativistic electrons $\dot{\gamma}_{\rm syn}$. %since the power of electrons radiating at frequency $\nu$ cannot exceed black-body. 
We thus modify the optically thin synchrotron cooling rate in eq.~(\ref{eq:gammadot_syn}) by a factor
\begin{equation} \label{eq:fssa}
f_{\rm ssa}(\gamma) \approx \frac{\rm radiated~power~at~\gamma}{\rm optically~thin~power~at~\gamma} \approx \frac{1 - e^{-\tau(\gamma)}}{\tau(\gamma)}
\end{equation}
where 
%$\tau(\gamma) = \alpha_\nu \left[ \nu=\nu_{\rm c}(\gamma) \right] R_{\rm n}$ 
%$\tau(\gamma) = \alpha_{\nu=\nu_{\rm c}} R_{\rm n}$
$\tau(\gamma) = \alpha_{\nu} \left( \nu_{\rm c} \left[ \gamma \right] \right) R_{\rm n}$
is an estimate of the synchrotron optical depth for an electron radiating at Lorentz factor $\gamma$.

Given the electron distribution function and magnetic field, we calculate the rotation measure through the nebula according to
\begin{align}
%&&{\rm RM} = \frac{e^{3}}{2\pi m_e^{2}c^{4}}\int n_e B_{\parallel} ds \approx  \frac{3e^{3}}{8\pi^{2} m_e^{2}c^{4}} \frac{N_{\rm e}B_{\rm n}}{R_{\rm n}^{2}}\left(\frac{\lambda}{R_{\rm n}}\right)^{1/2}\nonumber 
%{\rm RM} &= \frac{e^{3}}{2\pi m_e^{2}c^{4}}\int N_\gamma \frac{\ln \gamma}{2 \gamma^2} B_{\parallel} \, d\gamma ds
%\nonumber \\
%&\approx \frac{3e^{3}}{8\pi^{2} m_e^{2}c^{4}} \frac{B_{\rm n}}{R_{\rm n}^{2}}\left(\frac{\lambda}{R_{\rm n}}\right)^{1/2} \int N_\gamma \frac{\ln \gamma}{2 \gamma^2} \, d\gamma
{\rm RM} \approx \frac{e^{3}}{2\pi m_e^{2}c^{4}} \left(\frac{\lambda}{R_{\rm n}}\right)^{1/2} R_{\rm n} B_{\rm n} \int N_\gamma \frac{1}{\gamma^2} \, d\gamma
\label{eq:RM}
\end{align}
%where we have assumed a spacially constant electron density density $( \partial n_e / \partial \gamma ) = 3 N\gamma / 4\pi R_{\rm n}^3$,
where $\lambda \le R_{\rm n}$ quantifies the correlation length-scale of the magnetic field in the nebula,
and the $1/\gamma^2$ correction is an approximate interpolation between the non- and ultra-relativistic regimes \citep{Quataert&Gruzinov00b}.

%We additionally compute the synchrotron spectral energy density emitted by the population of relativistic electrons in the nebula, which in the optically thin regime is
%\begin{equation}
%L_\nu = \int \frac{2\pi e^3 B_{\rm n}}{\sqrt{3} \pi m_e c^2}F\left(\frac{\nu}{\nu_{\rm c}(\gamma)}\right) N_\gamma \, d\gamma .
%\end{equation}
%Here $\nu_{\rm c}(\gamma) = \gamma^2 eB / 2\pi m_e c$ is the cyclotron frequency of electrons with Lorentz factor $\gamma$, and $F(x) = x \int_{x}^{\infty} K_{5/3}(y) dy$ \citep[e.g.][]{Rybicki&Lightman}.
%Synchrotron self-absorption is treated in a simplified manner by taking the minimum of the above expression and
%\begin{equation}
%L_{\nu, {\rm sa}} \approx 4\pi R_{\rm n}^2 \frac{2 k T(\nu) \nu^2}{c^2} \approx \frac{8\pi}{3} R_{\rm n}^2 m_e \nu^2 \gamma(\nu)
%\end{equation}
%where the effective electron temperature $3 k T(\nu) = \gamma(\nu) m_e c^2$ is set by the Lorentz factor of electrons emitting at $\nu=\nu_{\rm c}$.

\section{Results}
\label{sec:results}
%Table~\ref{tab:models} lists the parameters for a few representative models which reasonably match the observed properties of the repeater, namely: the RM, the persistent radio flux and spectrum, and the inferred age. As discussed further in \S\ref{sec:analytic}, the problem is highly constrained mainly due to the difficulty in obtaining enough low Lorentz factor electrons to contribute to the high RM while simultaneously retaining enough high Lorentz fact electrons to produce the persistent radio flux. The existence of the solutions described below is therefore non-trivial.

We first describe some general features of the nebula evolution, before moving onto a few specific models found to reproduce most of the observed features of FRB 121102.
The electron energy distribution, $N_\gamma$, which at $t_0$ tracks the injected distribution peaking at $\gamma \sim \bar{\gamma}$ (eq.~\ref{eq:gammai}), rapidly evolves due to synchrotron cooling which extends the electron population to low Lorentz factors. The nebula's compactness and large radiation field however imply a high self-absorption frequency, below which synchrotron cooling is less efficient (eqs.~\ref{eq:gammadot_syn},\ref{eq:fssa}).  This effectively maintains the cooling frequency locked to the self-absorption frequency, creating a ``pile-up'' of electrons at that Lorentz factor, $\gamma_{\rm cool,syn}$.  Inverse Compton emission can also be important at early times ($\sim t_0$) in cooling electrons below the self-absorption frequency, however it quickly becomes negligible as the nebula expands. Throughout the remainder of the nebula's evolution, the self-absorption frequency decreases while $\gamma_{\rm cool,syn}$ rapidly increases, and the integrated number of low Lorentz factor electrons does not increase significantly (both because cooling is less effective and because, for our chosen values of $\alpha > 1$ in eq.~\ref{eq:Edot}, most particles are injected at $t \sim t_0$).
After this point, the distribution is governed by adiabatic cooling which sets $N_\gamma \propto \gamma^{-\alpha}$ for $\gamma \lesssim \bar{\gamma}$ with an exponential cutoff above (the exponential cutoff is due to the assumed Maxwellian injection distribution).

\begin{figure}
\centering
\includegraphics[width=0.5\textwidth]{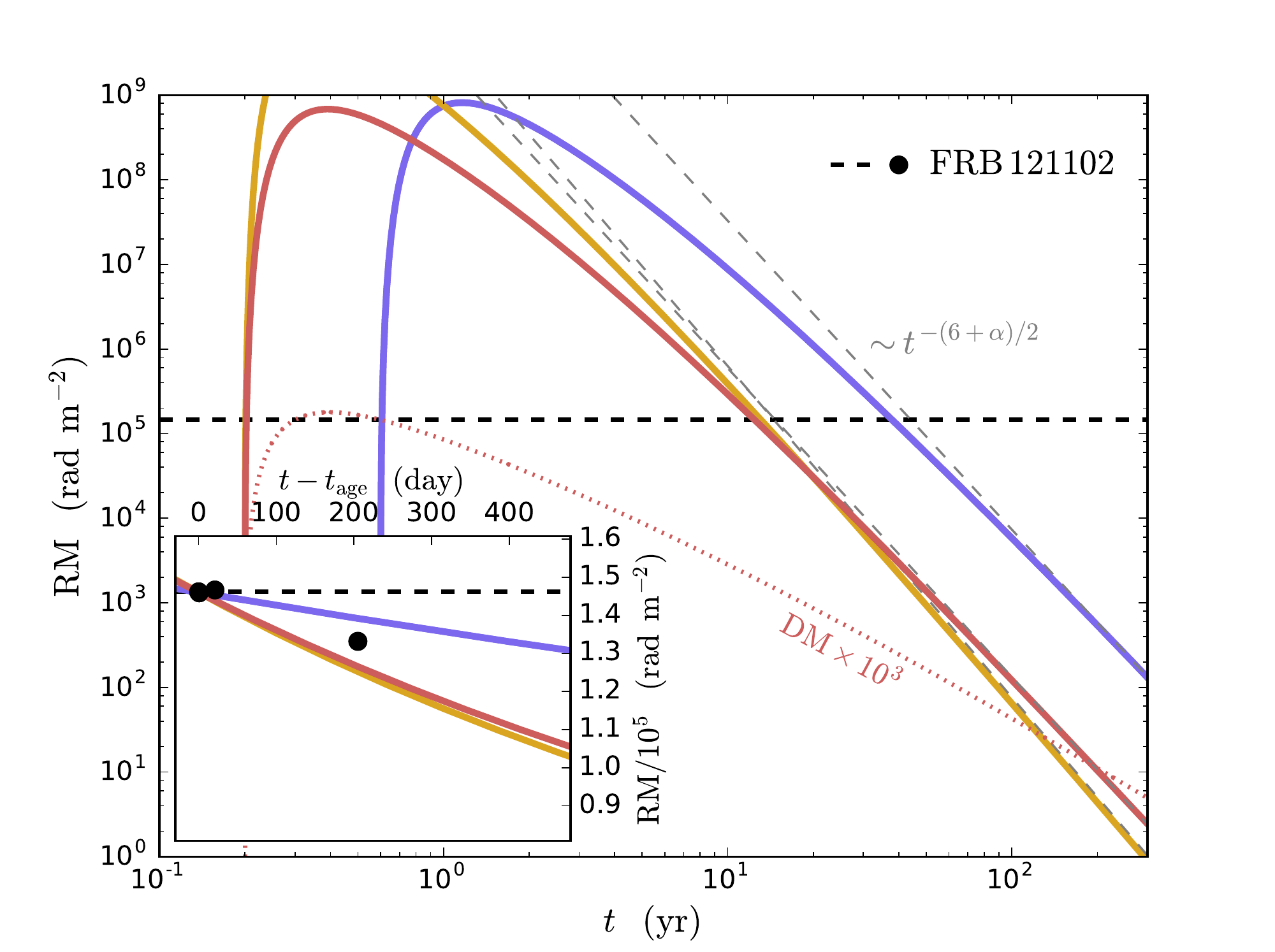}
\caption{Rotation measure as a function of time predicted by our representative models: A, B, and C (solid red, blue, and yellow curves, respectively). The horizontal black dashed curve shows the RM of FRB~121102 as first observed on MJD 57747 \citep{Michilli+18}, with later data points shown as black circles in the inset. The intersection between the red curve and this horizontal line imply an age of $t_{\rm age} \simeq 12.4$ yr for our fiducial model A. 
%($t_{\rm age} \simeq 37.8, 13.1 \, {\rm yr}$ for models D2, DR, respectively).
At times of interest the RM declines secularly, asymptoting to ${\rm RM} \propto t^{-(6+\alpha)/2}$ (dashed grey curves; eq.~\ref{eq:R17_constraint_RM}).
} \label{fig:RM}
\end{figure}

We take $E_{\rm B_{\star}} = 5 \times 10^{50} \, {\rm erg}$, $t_0 = 0.2 \, {\rm yr}$, $v_{\rm n} = 3 \times 10^8 \, {\rm cm \, s}^{-1}$, $\lambda = R_{\rm n}$, and $\alpha = 1.3$ for our fiducial model A and fix $\chi = 0.2 \, {\rm GeV} \approx \chi_{\min}$.
The RM as a function of time for this model is shown with a solid red curve in Fig.~\ref{fig:RM}. The dashed horizontal curve in this figure marks the observed ${\rm RM} = 1.46 \times 10^5 \, {\rm rad \, m}^{-2}$ for FRB~121102 as of late December 2016 (MJD 57747; \citealt{Michilli+18}). The intersection of the model-predicted RM with this curve defines the age of FRB~121102 within model A, yielding $t_{\rm age} \approx 12.4 \, {\rm yr}$.  The inset panel shows a zoom-in with the published RM data from \cite{Michilli+18} overlain as black circles. Our models generically predict a secular decline in RM with time, broadly consistent with the $\sim$ 10\% decline measured over a 7 month baseline.  The numerical models asymptote to the analytic estimate described in the next section (eq.~\ref{eq:R17_constraint_RM}; dashed-grey curves in Fig.~\ref{fig:RM}) which predicts an asymptotic power-law decline of ${\rm RM} \propto t^{-n}$ with $n = -(6+\alpha)/2 \approx -3.7$.  The actual slope at $t_{\rm age}$ is shallower than the analytic value, closer to $n \approx -3$ for model A.
%For $\alpha \sim 1-2$ this predicts a decay of $t^{-(3-4)}$, though 

%Figure~\ref{fig:RM} shows the RM as a function of time for our representative models. All models predict a rapid decrease in RM with time, falling off as \BM{$\sim t^{-(2-3)}$}.
Figure~\ref{fig:RM} also shows the dispersion measure (DM) due to the nebula for model A (dashed red curve; multiplied times one thousand), illustrating that it is extremely small 
%(in part due to the $1/\gamma$ suppression for relativistic electrons), 
and cannot contribute significantly to the inferred local $55 \lesssim {\rm DM} \lesssim 225 \, {\rm pc \, cm}^{-3}$ of FRB~121102 \citep{Tendulkar+17} when compared to other sources such as the surrounding supernova ejecta \citep[e.g.][]{Margalit+18}.

Two additional models, B and C, are also plotted in Fig.~\ref{fig:RM} (blue and yellow curves, respectively). As illustrated in Fig.~\ref{fig:timescales}, these models are chosen to yield:
three times larger $t_{\rm age}$ and same $R_{\rm n}(t_{\rm age})$ as model A, 
and
three times larger $R_{\rm n}(t_{\rm age})$ and same $t_{\rm age}$ as model A, respectively.
The following set of parameters were thus chosen: 
$E_{B_{\star}} = 5 \times 10^{50} \, {\rm erg}$, $t_0 = 0.6 \, {\rm yr}$, $v_{\rm n} = 10^8 \, {\rm cm \, s}^{-1}$, and $\alpha = 1.3$
for model B; and
$E_{\rm B_{\star}} = 4.9 \times 10^{51} \, {\rm erg}$, $t_0 = 0.2 \, {\rm yr}$, $v_{\rm n} = 9 \times 10^8 \, {\rm cm \, s}^{-1}$, and $\alpha = 1.83$
for model C,
resulting in $t_{\rm age} \simeq 37.8 \, {\rm yr}$ and $t_{\rm age} \simeq 13.1 \, {\rm yr}$ for the two models, respectively.

\begin{figure}
\centering
\includegraphics[width=0.5\textwidth]{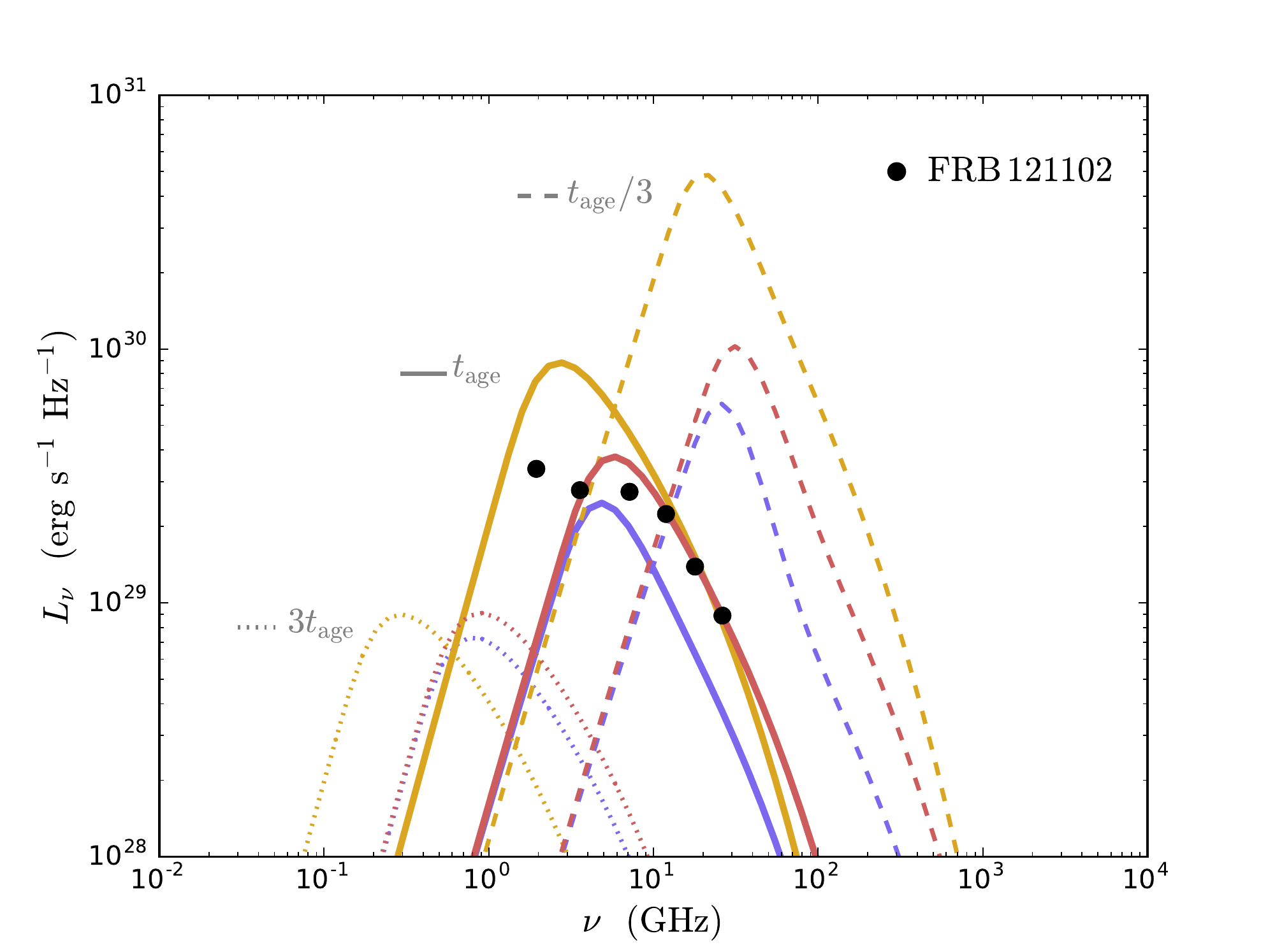}
\includegraphics[width=0.5\textwidth]{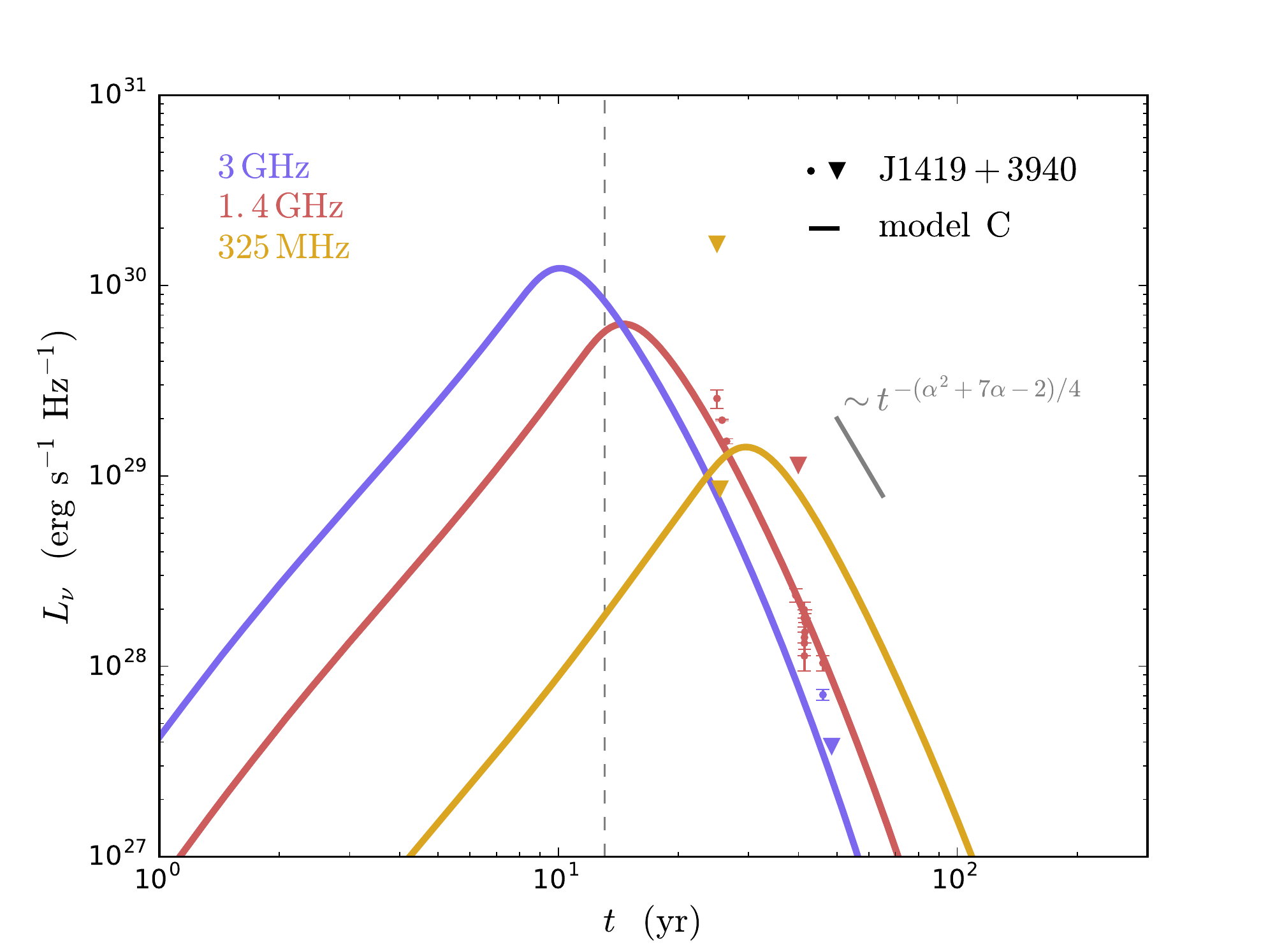}
\caption{Radio synchrotron emission from the nebula.   Top: spectral energy distribution at the observed epoch $t_{\rm age}$ (solid curves; models color coded as in Fig.~\ref{fig:RM}) and at $t/t_{\rm age} = 1/3, 3$ (dashed, dotted curves, respectively). Overlain in black circles is the persistent source associated with FRB~121102 \citep{Chatterjee+17}. The models predict a radio source broadly consistent FRB~121102 at an epoch when the model RM also fits the repeater (Fig.~\ref{fig:RM}).
Bottom: light-curve at $3 \, {\rm GHz}$, $1.4 \, {\rm GHz}$, and $325 \, {\rm MHz}$ (blue, red, and yellow curves, respectively) for model C.  The age of the source for this model is shown as a vertical dashed line.  The newly discovered radio source J1419+3940 hypothesized to be related in nature to the persistent radio source associated with FRB~121102 is plotted in comparison \citep{Law+18}. The model qualitatively agrees with the data, though no attempt was made to tune the model parameters for this application (beyond an arbitrary offset in the x-axis).}
\label{fig:luminosity}
\end{figure}

Radio light-curves and spectra for our models are shown in Fig.~\ref{fig:luminosity}. The top panel shows snapshots of the synchrotron spectrum at $t = t_{\rm age} - 100 \, {\rm d} \approx t_{\rm age}$ (solid curves) and at $t = t_{\rm age}/3$ and $t = 3 t_{\rm age}$ (dashed and dotted curves, respectively). The offset of $100 \, {\rm d}$  roughly corresponds to the time difference in the source frame between the first ${\rm RM}$ measurement (\citealt{Michilli+18}; to which we normalize $t=t_{\rm age}$) and the date at which the VLA spectrum of the persistent source was measured (\citealt{Chatterjee+17}; shown as black circles), but is not significant.
Our models produce a spectrum which is broadly consistent with the data, though an ``exact'' match is lacking.
In particular, models A and C exhibit a self-absorption break around $\nu \sim 5 \, {\rm GHz}$, in tension with the lowest frequency data point. It is possible that geometric effects extending the one-zone picture presented here may contribute to smooth the turnover around $\nu = \nu_{\rm ssa}$ (by effectively viewing multiple emitting regions with different optical depths) and alleviate this tension.  Alternatively, the heated electron spectrum may not be precisely that of a single-temperature relativistic Maxwellian.  A detailed exploration of such extensions is beyond the scope of this {\it Letter}.

The lower panel of Fig.~\ref{fig:luminosity} show the light-curve for model C at $3 \, {\rm GHz}$, $1.4 \, {\rm GHz}$, and $325 \, {\rm MHz}$ (blue, red, and yellow curves, respectively). The dashed vertical curve marks $t_{\rm age}$ for this model, for which the GHz band light-curve is near its peak. Additionally shown is data from the transient radio source J1419+3940 recently discovered by \cite{Law+18} in a small star-forming galaxy with remarkably similar properties to the host of FRB~121102 \citep{Ofek17}.  Although this event could represent the orphan radio afterglow of a long GRB, our modeling supplies some additional support for the magnetar nebula hypothesis.  More broadly, for our range of models, 
%at the present epoch, 
the light curve is predicted to decay as $F_{\nu} \propto t^{-m}$ with $m = -(\alpha^2+7\alpha-2)/4 \approx -3.5$ sufficiently late after the light-curve peak, as set by self-absorption. At times of interest ($\sim t_{\rm obs}$), our representative models are not yet within this regime at GHz frequencies (e.g. Fig.~\ref{fig:luminosity}) and the effective slope $m$ is smaller. Additionally, at late times an exponential cutoff due to the Maxwellian injection distribution steepens the slope significantly.

\subsection{Analytic Estimates}
\label{sec:analytic}

The model parameters needed to explain FRB~121102 are highly constrained by the requirements to reproduce the RM and its time derivative, while simultaneously obtaining the persistent source luminosity and spectrum.  Here we present analytic constraints on the nebula radius and energy based on the arguments above.

Accounting for adiabatic losses, the magnetic field in the nebula at time $t$ is a function of the energy injected within the last expansion timescale $\sim \dot{E}t = (\dot{E}t)_{50} 10^{50} \, {\rm erg}$,
\begin{equation} \label{eq:Bn}
B_{\rm n} \approx 0.24 \, {\rm G} \, \sigma_{-1}^{1/2} R_{17}^{-3/2} \left(\dot{E}t\right)_{50}^{1/2}
\propto t^{-(2+\alpha)/2},
\end{equation}
where $\sigma_{-1} \equiv \sigma/0.1$ and $R_{17} \equiv R_{\rm n}/10^{17}$ cm.
%and $(\dot{E}t)_{50} \equiv \dot{E}t / 10^{50} \, {\rm erg}$. 
The number of non-relativistic electrons in the nebula
\begin{equation} \label{eq:N0}
N_0 \approx 4 \times 10^{53} \chi_{0.2}^{-1} E_{50} ,
\end{equation}
is conversely set by the injection of energy at early times when cooling is efficient, $\sim E_{B_{\star}} \equiv E_{50}10^{50}{\rm erg}$, where $\chi_{0.2} \equiv \chi / 0.2 \, {\rm GeV}$ and we have assumed small magnetization, i.e. $(1+\sigma) \approx 1$.  The RM is proportional to the number of non-relativistic electrons (eq.~\ref{eq:N0}) and the magnetic field (eq.~\ref{eq:Bn}),
\begin{equation} \label{eq:R17_constraint_RM}
%R_{17} \approx 1.68 \sigma_{-1}^{1/7} \chi_{\min}^{-2/7} E_{0,50}^{2/7} \left(\dot{E}t\right)_{50}^{1/7} \left( \frac{\rm RM}{10^5 \, {\rm rad \, m}^{-2}}\right)^{-2/7} ,
{\rm RM}_{5} \approx 6 \times \sigma_{-1}^{1/2} \chi_{0.2}^{-1} E_{50} \left(\dot{E}t\right)_{50}^{1/2} R_{17}^{-7/2} \propto t^{-(6+\alpha)/2} ,
\end{equation}
where ${\rm RM}_{5} \equiv {\rm RM} / 10^5 \, {\rm rad \, m}^{-2}$.

The light-curve's temporal scaling can similarly be estimated. 
As the magnetic field dillutes (eq.~\ref{eq:Bn}), emission at frequency $\nu$ originates from higher Lorentz factor electrons, so that $\gamma(\nu) \propto t^{(2+\alpha)/4}$. 
As discussed in \S\ref{sec:results}, the electron distribution is frozen-in at late times, following $N_\gamma \propto \gamma^{-\alpha} t^{-(\alpha+2)}$ with an exponential cutoff above $\gamma \sim \bar{\gamma}$. We can therefore express the synchrotron luminosity as 
\begin{equation} \label{eq:Lnu}
L_\nu \approx \frac{4 \pi e^3}{m_{\rm e} c^2} R_{\rm n}^3 \gamma N_\gamma B_{\rm n} \propto \nu^{-(\alpha-1)/2} t^{-(\alpha^2+7\alpha-2)/4} .
\end{equation}
It is important to note that this scaling is only correct when both $\nu \gg \nu_{\rm ssa}(t)$ and $\gamma(\nu,t) \ll \bar{\gamma}$, 
the former of which does not hold
%which is not the case 
for GHz frequencies at $\sim t_{\rm obs}$ in our representative models.

%The synchrotron luminosity is a function of the number of recently injected relativistic electrons in the nebula, $\lesssim \dot{N} t$ (for fast cooling electrons $t_{\rm cool} < t$ and the relevant number of emitting electrons is $\dot{N} t_{\rm cool}$). Expressing the specific luminosity as 
%\begin{equation}
%L_\nu \approx \frac{2 e^3 B_{\rm n} N}{\sqrt{3} m_e c^2} \approx 3.8 \times 10^{29} \, {\rm erg \, s}^{-1} \, {\rm Hz}^{-1} \, \left(\frac{B_{\rm n}}{0.24 \, {\rm G}}\right) \left(\frac{N}{10^{52}}\right)
%\end{equation}
%we find from the above constraint that
%\begin{equation} \label{eq:R17_constraint_Lnu}
%R_{17} \lesssim 12.7 \sigma_{-1}^{1/3} \chi_{\min}^{-2/3} \left(\dot{E}t\right)_{50} \left( \frac{L_\nu}{3.3 \times 10^{29} \, {\rm erg \, s}^{-1} \, {\rm Hz}^{-1}}\right)^{-2/3}. 
%\end{equation}

The lack of clear self-absorption in the persistent source's spectrum indicates that $\nu_{\rm ssa} \lesssim \nu_{\rm obs}$. This condition is equivalent to the requirement that the synchrotron specific intensity at $\nu$ not exceed that of a black-body with an effective temperature $k T = \gamma(\nu) m_{\rm e} c^2 / 3$, where $\gamma (\nu)$ is 
%given by eq.~(\ref{eq:nuc}) with 
the electron Lorentz factor emitting at
$\nu \simeq 0.29 \nu_{\rm c}(\gamma)$. In the Rayleigh-Jeans regime this implies that $L_\nu \lesssim 4\pi^2 R_{\rm n}^2 \times 2 k T \nu^2 / c^2$ and leads to the constraint
\begin{equation} \label{eq:R17_constraint_ssa}
%R_{17} \gtrsim 2.48 \sigma_{-1}^{1/11} \left(\dot{E}t\right)_{50}^{1/11} \left( \frac{L_\nu}{3.3 \times 10^{29} \, {\rm erg \, s}^{-1} \, {\rm Hz}^{-1}} \right)^{4/11} \left( \frac{\nu}{1.9 \, {\rm GHz}} \right)^{-10/11} .
R_{17} \gtrsim 0.46 \sigma_{-1}^{1/11} \left(\dot{E}t\right)_{50}^{1/11} L_{\nu,{\rm obs}}^{4/11} \nu_{\rm obs}^{-10/11}
\end{equation}
for $L_{\nu,{\rm obs}} \simeq 2.7 \times 10^{29} \, {\rm erg \, s}^{-1} \, {\rm Hz}^{-1}$ and $\nu_{\rm obs} \simeq 7.2 \, {\rm GHz}$ corresponding to the redshift-corrected luminosity of the persistent source observed at $6 \, {\rm GHz}$ \citep{Chatterjee+17}.
This criterion, plotted in Fig.~\ref{fig:timescales}, is only weakly dependent on the model parameters $\dot{E}t$, $\sigma$, but becomes more stringent if the $1.6 \, {\rm GHz}$ frequency VLA data is assumed to be unabsorbed as well, yielding $R_{17} \gtrsim 1.62$.

Combining eqs.~(\ref{eq:R17_constraint_RM},\ref{eq:R17_constraint_ssa}) and requiring ${\rm RM}_5 \simeq 1.46$ to fit the observed ${\rm RM}_{\rm obs}$ \citep{Michilli+18}, 
and additionally imposing $\dot{E}t < E_{\rm B_{\star}}$ as appropriate for a decreasing energy injection rate (eq.~\ref{eq:Edot}), we find that
\begin{equation}
E_{\rm B_{\star}}> 4 \times 10^{48} \, {\rm erg} \,  \sigma_{-1}^{-2/13} \chi_{0.2}^{11/13} 
L_{\nu,{\rm obs}}^{21/26} \nu_{\rm obs}^{-35/13} {\rm RM}_{\rm obs}^{11/13}
\label{eq:Econstraint}
\end{equation}
is required to satisfy both the RM and self-absorption constraints. 
Eq.~\ref{eq:Econstraint} is a strong function of frequency, such that if one assumes that the observer-frame self-absorption frequency is instead below $1.6 \, {\rm GHz}$, a more stringent constraint of $E_{\rm B_{\star}} > 1.7 \times 10^{50} \, {\rm erg}$ is obtained.

\section{Discussion and Implications}

The radio flux and RM contribution of an expanding magnetized electron-ion nebula, inflated behind the supernova ejecta by a flaring young magnetar, are consistent with the observed properties of the repeating burster FRB 121102 for source ages $t_{\rm age} \sim 10-40$ yr consistent with a variety of other observational constraints (Fig.~\ref{fig:timescales}).  Our model predicts the presence of a self-absorption turnover in the spectral energy distribution that should be observable with low frequency observations.  Our detailed calculations broadly follow the scenario outlined by \citet{Beloborodov17}.  However, we find that non-thermal particle acceleration or sustained heating of the nebular electrons
(e.g. \citealt{Yang+16})
is not required.  Thermal heating of electrons at the termination shock, for wind baryon-loading $\chi \sim \chi_{\rm min}$ inferred from giant Galactic flares \citep{Granot+06} and the natural scale set by the gravitational potential of a neutron star, is sufficient to explain all the available observations.

We  predict approximate power-law decays of the RM $\propto t^{-(6+\alpha)/2}$ and persistent source flux $F_{\nu} \propto t^{-(\alpha^2+7\alpha-2)/4}$ (eqs.~\ref{eq:R17_constraint_RM},\ref{eq:Lnu}), where $\alpha \gtrsim 1$ sets the magnetar energy injection rate, $\dot{E} \propto t^{-\alpha}$ (eq.~\ref{eq:Edot}).
For the RM, this is an asymptotic scaling and the slope of RM versus $t$ is generally somewhat shallower, while the analytic result for $F_{\nu}$ is significantly modified at times of interest $\sim t_{\rm obs}$ due to self-absorption (decreasing the effective slope dramatically) and by an exponential cutoff at very late times (increasing it).
Importantly, in both cases our models describe the secular trend averaged over long baselines, as the turbulent environment of the nebula (and of the ISM of the host galaxy or Milky Way) could  produce shorter timescale fluctuations. 
Our predicted long-baseline secular evolution is distinct from the stochastic or periodic RM evolution implied by models attributing the RM to the environment near a galactic nucleus \citep[e.g.][]{Thompson17,Zhang18}, and thus provides a possible way to distinguish such models.

Although FRB\,121102 can be understood in the magnetar picture, the model does place stringent requirements on the source properties.  The total energy of the magnetar likely must obey $E_{B_\star} \gtrsim  10^{50}$ erg, requiring a large interior magnetic field strength, $B_\star \gtrsim 2 \times 10^{16} \, {\rm G}$ (eq.~\ref{eq:EB}).  While seemingly extreme, such a magnetic energy still represents less than a few percent of the rotational energy present in a millisecond magnetar, the latter being a requirement for powering a GRB or superluminous supernova.  Such a strong field may also be required for magnetic flux to emerge from the magnetar on the requisite short timescale of decades (eq.~\ref{eq:tmag}). 

To reproduce the RM of FRB~121102, the radial component of the nebular magnetic field must possess a coherence length comparable to the nebula size ($\lambda \sim R_{\rm n}$ in eq.~\ref{eq:RM}).  
%It is somewhat unclear how this would arise physically if the flares feeding the nebula are radomly-oriented.  
Such an ordered field may also be supported {\it empirically} by the  $\sim$constant direction of the polarization vector of the bursts from FRB~121102 over several months \citep{Michilli+18}. If FRBs originate from the forward shock generated as flare ejecta collide with the magnetar wind \citep{Beloborodov17}, then this indicates that the upstream magnetic field of the wind itself is fixed in its direction over many flare timescales. 

One way such a large coherence length might be established
is through self-organization of an initially random magnetic field due to an inverse energy cascade in relativistic magnetohydrodynamic turbulence \citep{Zrake14}.
In this scenario, the RM may randomly reverse sign over an eddy turnover timescale $\sim R_{\rm n}/v_{\rm A} \sim 4 \, {\rm months} \, R_{17} \sigma_{-1}^{-1/2}$, where $v_{\rm A} = \sigma c$ is the Alfven speed. The secular decline implied by our model (eq.~\ref{eq:RM}) should then be interpreted as a long baseline envelope of $\vert {\rm RM} \vert$.

Another possibility is that rotation of the magnetar plays a role in setting the magnetic field orientation.
%a preferred direction. If the rotation period is is faster than the timescale over which an outflow is launched from the magnetar surface, then the magnetic field may ``wrap around'' the outflow }
Indeed, a toroidal field perpendicular to the rotation axis is a general feature of pulsar winds.  On the other hand, the build-up of too large an ordered field in the nebula could lead to non-axisymmetric kink instabilities \citep{Begelman98} and associated magnetic dissipation that regulates to an ordered component with $\sigma \lesssim 0.1$ (e.g.~\citealt{Porth+13}).  Such instabilities could also play a role in generating the necessary radial component of the field needed for the RM.

Our representative models have been chosen by hand, with no attempt to rigorously fit the data. Given the number of constraints imposed on the model and its relative simplicity, it is thus non-trivial that we have been able to find reasonable parameters which produce both RM and $L_\nu$ at a given epoch to within an order of magnitude, while also satisfying all other observational constraints (Fig.~\ref{fig:timescales}).  
As one example, we found that $\dot{E} = {\rm constant}$ ($\alpha = 0$) models cannot reproduce the observations, because the number of electrons injected at early times is too low for values of $\dot{E}$ which continue to power a sufficient radio luminosity at the source's age, resulting in the RM being underproduced.

Our finding that FRB~121102 requires an energy injection rate $\dot{E} \propto t^{-\alpha}$ with $\alpha \gtrsim 1$, has potential implications for its FRB activity.  This implies that either the rate of FRB activity will slow down, or that flares will on average become less energetic, over a timescale of decades. 
%Fig.~X shows the RM as a function of cumulative energy release with time for our best-fit models {\bf BDM: new figure here?}.  
%Assuming energy release tracks FRB activity, this can be taken as a prediction for the range of RM from a wider range of RM, e.g. assuming even currently non-repeating FRBs are also flaring magnetars.  
Assuming energy release tracks FRB activity, we can make a prediction for the range of RM for a population of FRB sources.
%, e.g. assuming even currently non-repeating FRBs are also flaring magnetars.  
Under the assumption that all (even currently non-repeating) FRBs are similar flaring magnetars, and that FRBs follow the release of magnetic energy, we can estimate the probability of detecting an FRB at a given RM. Using our analytic prediction for the dependence of RM on time (eq.~\ref{eq:R17_constraint_RM}) along with $E(t)$ from eq.~\ref{eq:Edot}, we find that $(dE/d{\rm RM}) {\rm RM} \propto {\rm RM}^{2(\alpha-1)/(6+\alpha)}$.  For the range of $\alpha$ adopted in our representative models, this implies a relatively ``flat'' distribution, e.g. $(dE/d{\rm RM}) {\rm RM} \propto {\rm RM}^{0.08}$ for our fiducial $\alpha=1.3$.  Although the RM is highest early in the nebula history when $\dot{E}$ is large, sufficient energy is released at later times that many sources should be detected once the RM has dropped to much lower values.

Consistent with such a distribution, a few FRBs other than FRB 121102 have measured RM values, ranging from small values $\lesssim 30 \, {\rm rad \, m}^{-2}$ consistent with the Galactic contribution (\citealt{Ravi+16,Petroff+17}) to higher values still less than in FRB~121102 (\citealt{Masui+15}).  Many FRBs with zero measured linear polarization could in fact have similarly high RM to FRB 121102, due to artificial depolarization caused if the observations are taken with insufficient frequency resolution \citep{Michilli+18}.

\bigskip

We thank Jonathan Zrake for helpful discussions.
%This research benefited from interactions with X that were funded by the Gordon and Betty Moore Foundation through Grant GBMF5076
This research benefited from interactions at the ZTF Theory Network Meeting, funded by the Gordon and Betty Moore Foundation through Grant GBMF5076.  B.D.M. and B.M. acknowledge support from NASA through the Astrophysics Research Program (grant number NNX16AB30G). 
Support for this work was provided by NASA through the NASA Hubble Fellowship grant \#HST-HF2-51412.001-A awarded by the Space Telescope Science Institute, which is operated by the Association of Universities for Research in Astronomy, Inc., for NASA, under contract NAS5-26555.

\bibliographystyle{yahapj}
\bibliography{refs}

%%%%%%%%%%%%%%%%%%%%%%%%%%%%%%%%%%%%%%%%%%%%%%%%%%

%% This command is needed to show the entire author+affilation list when
%% the collaboration and author truncation commands are used.  It has to
%% go at the end of the manuscript.
%\allauthors

%% Include this line if you are using the \added, \replaced, \deleted
%% commands to see a summary list of all changes at the end of the article.
%\listofchanges

\end{document}